\documentclass[10pt]{iopart}

\newcommand{\Rm}{\mathbb{R}}
\newcommand{\Cm}{\mathbb{C}}

\newcommand{\Sm}{\mathbb{S}}
\newcommand{\be}{\[}
\newcommand{\ee}{\]}
\newcommand{\ba}{\begin{eqnarray*}}
\newcommand{\ea}{\end{eqnarray*}}
\newcommand{\va}{\varphi}

\newcommand{\pp}{\partial}
\newcommand{\vv}[1]{\boldsymbol{\mathrm{#1}}}
\newcommand{\hvv}[1]{\boldsymbol{\hat{\mathrm{#1}}}}
\newcommand{\uv}{\boldsymbol{{\hat{\mathrm{s}}}}}
\newcommand{\uvk}{\boldsymbol{\hat{\mathrm{k}}}}

\newcommand{\pint}{\:\mathcal{P}\!\!\int}
\newcommand{\rrf}[1]{\mathop{\mathcal{R}_{{#1}}}}
\newcommand{\irrf}[1]{\mathop{\mathcal{R}_{{#1}}^{-1}}}

\usepackage{iopams}
\usepackage{amsthm,amssymb,mathrsfs}
\usepackage{graphicx}

\usepackage{color}

\newtheorem{thm}{Theorem}[section]
\newtheorem{lem}[thm]{Lemma}

\theoremstyle{remark}\newtheorem{rmk}[thm]{Remark}

\begin{document}

\title[]{Analytical discrete ordinates for the three-dimensional radiative radiative equation in the half space}

\author{Manabu Machida$^{1,2}$}
\address{$^1$ Institute for Medical Photonics Research,
Hamamatsu University School of Medicine,
Hamamatsu 431-3192, Japan}
\address{$^2$ JST, PRESTO, Kawaguchi, Saitama 332-0012, Japan}
\ead{machida@hama-med.ac.jp}

\begin{abstract}
In one-dimensional transport theory, the method of analytical discrete ordinates (ADO) is known to be a concise and fast numerical scheme to solve the radiative transport equation. However, the extension of ADO to three dimensions has been restricted to the case of isotropic scattering. In this paper, we extend ADO to the case of the three-dimensional radiative transport equation with anisotropic scattering. This extension is made possible by the technique of rotated reference frames. The radiative transport equation with constant absorption and scattering coefficients in the half space is solved by the extended ADO in three dimensions. Moreover, we show that even an analytical solution in the half space is obtained by extending the singular eigenfunction approach with rotated reference frames.
\end{abstract}

\maketitle

\section{Introduction}
\label{intro}

The radiative transport equation is a governing equation for transport phenomena which are characterized by scattering and absorption \cite{Case-Zweifel,Chandrasekhar50}. Indeed, the radiative transport equation is the linear Boltzmann equation. A typical example of such transport phenomena is the propagation of near-infrared light in biological tissue \cite{Carminati-Schotland}. Since the radiative transport equation has an integral term for the angular variable, solving the equation is difficult not only analytically but even numerically. Although Monte Carlo simulation, finite difference schemes, and finite element methods are often employed for the numerical calculation of the radiative transport equation, different efforts have been made about efficient numerical methods for the radiative transport equation.

This paper considers the radiative transport equation with constant coefficients in the half space with the discretization of the integral term by discrete ordinates. This situation was considered in \cite{Kim04}, in which the plane-wave decomposition \cite{Kim-Keller03} was employed and the three-dimensional problem reduces to a one-dimensional problem by the Fourier transform. Although the method has been proven to work well in the half space \cite{Kim04} and in the slab geometry \cite{Kim-Moscoso04}, separation constants or eigenvalues must be computed for every Fourier vector. In this paper, the plane-wave decomposition and separation of variables are used, but we show that only one diagonalization is enough to obtain the separation constant.

We will solve the three-dimensional radiative transport equation by extending the method of analytical discrete ordinates (ADO) to three dimensions. This was done in \cite{Machida-Das22} for an infinite medium. In this paper, the half space is considered.

In one-dimensional transport theory, ADO employs separation of variables for equations with discrete ordinates \cite{Barichello11,Barichello-Garcia-Siewert00,Barichello-Siewert99a,Barichello-Siewert02}. Although the searchlight problem was solved by ADO \cite{Barichello-Siewert00}, the scattering phase function had to be a constant. In one dimension, a more general scattering phase function with polar and azimuthal angles was considered for ADO \cite{Garcia-Siewert10}.

The rest of the paper is organized as follows. In Sec.~\ref{drte}, the three-dimensional radiative transport equation with discrete ordinates is introduced. In Sec.~\ref{dmodes}, eigenmodes are obtained. The energy density is considered in Sec.~\ref{dIspart}. Section \ref{dnum} is devoted to numerical calculation. Concluding remarks are given in Sec.~\ref{concl}. Finally, in \ref{Wignerpyramid} and \ref{singular}, analytically continued Wigner's $d$-matrices and singular eigenfunctions in three dimensions are described, respectively.

\section{Radiative transport equation with discrete ordinates}
\label{drte}

We consider the radiative transport equation in the half space $\Omega\subset\Rm^3$, i.e.,
\be
\Omega=\left\{\vv{r}\in\Rm^3;\,\vv{\rho}\in\Rm^2,\,z\in(0,\infty)\right\},
\ee
where $\vv{r}={^t}(\vv{\rho},z)$, $\vv{\rho}={^t}(x,y)$. Let $\pp\Omega$ denote the boundary at $z=0$. Let $\mu_a,\mu_s$ be the absorption and scattering coefficients, respectively. We assume that $\mu_a$ and $\mu_s$ are positive constants. The scattering phase function $p(\uv,\uv')$ is modeled as
\begin{equation}\fl
p(\uv,\uv')=
\frac{1}{4\pi}\sum_{l=0}^{l_{\rm max}}(2l+1){\rm g}^lP_l(\uv\cdot\uv')=
\sum_{l=0}^{l_{\rm max}}\sum_{m=-l}^l{\rm g}^lY_{lm}(\uv)Y_{lm}^*(\uv'),
\quad\uv,\uv'\in\Sm^2
\label{phasefunc}
\end{equation}
with $l_{\rm max}\ge0$, Legendre polynomials $P_l$, and spherical harmonics $Y_{lm}$. Spherical harmonics are defined as
\be\fl
Y_{lm}(\uv)=
\sqrt{\frac{2l+1}{4\pi}\frac{(l-m)!}{(l+m)!}}P_l^m(\mu)e^{im\va},
\quad\uv\in\Sm^2,\quad-1\le\mu\le1,\quad 0\le\va<2\pi,
\ee
where $P_l^m(\mu)$ are associated Legendre polynomials, and $\va,\mu$ are the azimuthal angle and the cosine of the polar angle of $\uv$. The constant ${\rm g}\in[0,1)$ is the anisotropic factor.

Let us discretize $\mu$. With the Gauss-Legendre quadrature, we let $\mu_i,w_i$ ($i=1,\dots,N$) be abscissas and weights. They can be computed by the Golub-Welsch algorithm \cite{GW}. We label $\mu_i$ ($i=1,\dots,N$) such that $0<\mu_1<\mu_2<\cdots<\mu_N<1$ and $-1<\mu_{2N}<\cdots<\mu_{N+2}<\mu_{N+1}<0$ ($\mu_{N+i}=-\mu_i$, $1\le i\le N$). For sufficiently large $N$, we have
\be
\sum_{i=1}^{2N}\int_0^{2\pi}p(\uv_i,\uv_{i'})\,d\va=1,\quad
\sum_{i'=1}^{2N}\int_0^{2\pi}(\uv_i\cdot\uv_{i'})p(\uv_i,\uv_{i'})\,d\va'={\rm g}.
\ee
The unit vector $\uv_i$ is introduced as
\be\fl
\uv_i=\left(\begin{array}{c}\vv{\omega}_i\\ \mu_i\end{array}\right),
\quad
\vv{\omega}_i=\left(\begin{array}{c}
\sqrt{1-\mu_i^2}\cos\va \\ \sqrt{1-\mu_i^2}\sin\va
\end{array}\right),\quad
i=1,\dots,2N,\quad 0\le\va<2\pi.
\ee

Let $I(\vv{r},\uv_i)$ be the specific intensity at position $\vv{r}\in\Rm^3$ in direction $\uv_i$. The specific intensity obeys the following radiative transport equation.
\begin{equation}
\cases{
\left(\uv_i\cdot\nabla+\mu_t\right)I(\vv{r},\uv_i)=
\mu_s\sum_{i'=1}^{2N}w_{i'}\int_0^{2\pi}p(\uv_i,\uv_{i'})I(\vv{r},\uv_{i'})\,d\va',
&
\\
\qquad\vv{r}\in\Omega,\quad i=1,\dots,2N,\quad 0\le\va<2\pi,&
\\
I(\vv{r},\uv_i)=g(\vv{\rho},\uv_i),&
\\
\qquad\vv{r}\in\pp\Omega,\quad i=1,\dots,N,\quad 0\le\va<2\pi,&
}
\label{intro:rte0}
\end{equation}
with $I\to0$ as $|\vv{r}|\to\infty$. Here, $\nabla=\pp/\pp\vv{r}$, and $\mu_t=\mu_a+\mu_s$. Let $\mu_{i_0},\va_0$ be angles for a unit vector $\uv_{i_0}$. Let $\vv{\omega}_{i_0}$ be the two-dimensional vector for  $\mu_{i_0},\va_0$. We give the source term as
\begin{equation}
g(\vv{\rho},\uv_i)=\delta(\vv{\rho})\delta_{ii_0}\delta(\va-\va_0),
\label{sourceg}
\end{equation}
where $\delta(\vv{\rho})=\delta(x)\delta(y)$ and $\delta(\va-\va_0)$ denote Dirac's delta functions, and $\delta_{ii_0}$ is the Kronecker delta.

Let us divide both sides of (\ref{intro:rte0}) by $\mu_t$ and introduce dimensionless variables: $\vv{r}_*=\mu_t\vv{r}$, $\vv{\rho}_*=\mu_t\vv{\rho}$, $z_*=\mu_tz$, $\nabla_*=\pp/\pp\vv{r}_*$, $I_*(\vv{r}_*,\uv)=I(\vv{r},\uv)$, and $g_*(\vv{\rho}_*,\uv)=g(\vv{\rho},\uv)$. We note that
\be
g_*(\vv{\rho}_*,\uv_i)=
\mu_t^2\delta(\vv{\rho}_*)\delta_{ii_0}\delta(\va-\va_0).
\ee
By dropping the subscript $*$, the radiative transport equation is written as
\begin{equation}
\cases{
\left(\uv_i\cdot\nabla+1\right)I(\vv{r},\uv_i)=
\varpi\sum_{i'=1}^{2N}w_{i'}\int_0^{2\pi}p(\uv_i,\uv_{i'})I(\vv{r},\uv_{i'})\,d\va',
&
\\
\qquad\vv{r}\in\Omega,\quad i=1,\dots,2N,\quad 0\le\va<2\pi,&
\\
I(\vv{r},\uv_i)=g(\vv{\rho},\uv_i),&
\\
\qquad\vv{r}\in\pp\Omega,\quad i=1,\dots,N,\quad 0\le\va<2\pi&
}
\label{intro:rte}
\end{equation}
with $I\to0$ as $|\vv{r}|\to\infty$. Here, $\varpi=\mu_s/\mu_t\in(0,1)$ is the albedo for single scattering.

Let us consider the Fourier transform
\begin{equation}\fl
\widetilde{I}(\vv{q},z,\uv)=
\int_{\Rm^2}e^{-i\vv{q}\cdot\vv{\rho}}I(\vv{r},\uv)\,d\vv{\rho},
\quad
I(\vv{r},\uv)=\frac{1}{(2\pi)^2}\int_{\Rm^2}e^{i\vv{q}\cdot\vv{\rho}}
\widetilde{I}(\vv{q},z,\uv)\,d\vv{q}.
\label{asol:fourier}
\end{equation}
We write $\vv{q}\in\Rm^2$ as
\be
\vv{q}=q\left(\begin{array}{c}\cos\va_{\vv{q}}\\\sin\va_{\vv{q}}\end{array}\right),\quad
q\ge0,\quad0\le\va_{\vv{q}}<2\pi.
\ee
The Fourier transform $\widetilde{g}(\vv{q},\uv)$ is written as
\be
\widetilde{g}(\vv{q},\uv_i)=\mu_t^2\delta_{ii_0}\delta(\va-\va_0).
\ee

In the Fourier space, $\widetilde{I}(\vv{q},z,\uv_i)$ is obtained as the solution to the following equation.
\begin{equation}
\cases{
\left(\mu_i\frac{\pp}{\pp z}+1+i\vv{\omega}_i\cdot\vv{q}\right)
\widetilde{I}=
\varpi\sum_{i'=1}^{2N}w_{i'}\int_0^{2\pi}p(\uv_i,\uv_{i'})\widetilde{I}(\vv{r},\uv_{i'})\,d\va',
\\
\qquad z\in(0,\infty),\; i=1,\dots,2N,\; 0\le\va<2\pi,
\\
\widetilde{I}=\widetilde{g}(\vv{q},\uv_i),
\quad z=0,\; i=1,\dots,N,\; 0\le\va<2\pi.
}
\label{intro:I2Feq}
\end{equation}
We note that
\be
\vv{\omega}_i\cdot\vv{q}=q\sqrt{1-\mu_i^2}\cos(\va-\va_{\vv{q}}).
\ee
Let us define
\begin{equation}
\omega_i^0(q,\va)=q\sqrt{1-\mu_i^2}\cos\va,
\label{omegai0}
\end{equation}
where $q=|\vv{q}|$.

\section{Eigenmodes}
\label{dmodes}

We seek solutions of the form of plane-wave expansion \cite{Kaper69,Kim04,Kim-Keller03}.  We introduce $\nu\in\Rm$ and $\vv{q}\in\Rm^2$, and define vector $\vv{k}\in\Cm^3$ as \cite{Markel04}
\begin{equation}
\vv{k}=\frac{1}{\nu}\uvk,\quad
\uvk=\left(\begin{array}{c}-i\nu\vv{q} \\ \hat{k}_z(\nu q)\end{array}\right),\quad
\hat{k}_z(\nu q)=\sqrt{1+(\nu q)^2}.
\label{defveck}
\end{equation}
We note that $\uvk$ is a unit vector in the sense of $\uvk\cdot\uvk=1$. We have
\begin{equation}\fl
\uv_i\cdot\uvk=
-i\nu\vv{\omega}_i\cdot\vv{q}+\hat{k}_z(\nu q)\mu_i=
-i\nu q\sqrt{1-\mu_i^2}\cos(\va-\va_{\vv{q}})+\hat{k}_z(\nu q)\mu_i.
\label{modes:sk}
\end{equation}

Let us consider the homogeneous equation below.
\begin{equation}\fl
\left(\mu_i\frac{\pp}{\pp z}+1+i\vv{\omega}_i\cdot\vv{q}\right)
\check{I}(\vv{q},z,\uv_i)=
\varpi\sum_{i'=1}^{2N}w_{i'}\int_0^{2\pi}p(\uv_i,\uv_{i'})\check{I}(\vv{q},z,\uv_{i'})\,d\va'
\label{modes:homoeq}
\end{equation}
for $\vv{r}\in\Rm^3$, $\uv_i\in\Sm^2$. Let $\rrf{\uvk}$ be the operator which rotates the reference frame in such a way that the $z$-axis is rotated to the direction of $\uvk$ \cite{Machida15,Markel04}. In particular, noticing $\mu_i=\uv_i\cdot\hvv{z}$, where $\hvv{z}={^t}(0,0,1)$, we can write
\be
\rrf{\uvk}\mu_i=\uv_i\cdot\uvk.
\ee
We assume the following separated solution.
\begin{equation}
\check{I}(\vv{q},z,\uv_i)=
\rrf{\uvk(\nu,\vv{q})}\Phi_{\nu}^m(\uv_i)e^{-\hat{k}_z(\nu q)z/\nu},
\label{modes:separatedI}
\end{equation}
where
\begin{eqnarray}
\Phi_{\nu}^m(\uv_i)
&=
\Phi_{\nu}^m(\mu_i,\va)
\nonumber \\
&=
\phi^m(\nu,\mu_i)\left(1-\mu_i^2\right)^{|m|/2}e^{im\va}.
\label{modes:Phiphi}
\end{eqnarray}
The function $\phi^m(\nu,\mu_i)$ is normalized as
\begin{equation}
\sum_{i=1}^{2N}w_i\phi^m(\nu,\mu_i)\left(1-\mu_i^2\right)^{|m|}=1.
\label{normcond}
\end{equation}
In the laboratory frame ($\uvk=\hvv{z}$), (\ref{modes:separatedI}) reduces to the eigenvectors of the original ADO.

Since an integral over all angles does not change when the reference frame is rotated, the normalization condition (\ref{normcond}) implies
\be
\frac{1}{2\pi}\sum_{i=1}^{2N}w_i\int_0^{2\pi}\phi^m(\nu,\uv_i\cdot\uvk)
\left[1-(\uv_i\cdot\uvk)^2\right]^{|m|}\,d\va\approx1.
\ee
That is, the left-hand side of the above equation is numerically $1$, even if it is not exactly $1$ due to the discretization of $\mu$.

Let $\theta_{\uvk},\va_{\uvk}$ be the polar and azimuthal angles of $\uvk=\uvk(\nu,\vv{q})$. We obtain \cite{Machida14}
\be
\va_{\uvk}=\cases{
\va_{\vv{q}}+\pi&for $\nu>0$,
\\
\va_{\vv{q}}&for $\nu<0$,
}
\ee
and
\be
d_{m'm}^l(\theta_{\uvk})=d_{m'm}^l[i\tau(\nu q)].
\ee
See \ref{Wignerpyramid} for analytically continued Wigner's $d$-matrices $d_{m'm}^l(\theta_{\uvk})$. We note that
\be
\rrf{\uvk}Y_{lm}(\uv)=\sum_{m'=-l}^le^{-im'\va_{\uvk}}d_{m'm}^l(\theta_{\uvk})Y_{lm'}(\uv).
\ee
Using $Y_{lm}^*(\uv)=(-1)^mY_{l,-m}(\uv)$, we have
\ba
\rrf{\uvk}Y_{lm}^*(\uv)
&=
(-1)^m\sum_{m'=-l}^le^{-im'\va_{\uvk}}d_{m',-m}^l(\theta_{\uvk})Y_{lm'}(\uv)
\\
&=
\sum_{m'=-l}^le^{im'\va_{\uvk}}d_{m'm}^l(\theta_{\uvk})Y_{lm'}^*(\uv).
\ea

\begin{rmk}
The idea of rotated reference frames first appeared in the context of the $P_L$ approximation \cite{Kobayashi77}. Later, the rotation was independently developed and an efficient numerical algorithm was devised \cite{Markel04,Panasyuk06}.
\end{rmk}

By substitution we have
\begin{equation}\fl
\left(1-\frac{\uv_i\cdot\uvk(\nu,\vv{q})}{\nu}\right)
\rrf{\uvk(\nu,\vv{q})}\Phi_{\nu}^m(\uv_i)=
\varpi\sum_{i'=1}^{2N}w_{i'}\int_0^{2\pi}p(\uv_i,\uv_{i'})
\rrf{\uvk(\nu,\vv{q})}\Phi_{\nu}^m(\uv_{i'})\,d\va'.
\label{modes:phieq}
\end{equation}
We note that $p(\uv_i,\uv_{i'})$ is invariant under rotation because it depends only on $\uv_i\cdot\uv_{i'}$. Since \cite{Varshalovich}
\be
\sum_{m=-l}^l(-1)^{m+m''}d_{m'm}^l(\theta_{\uvk})d_{mm''}^l(\theta_{\uvk})=
\delta_{m'm''},
\ee
we have
\ba
\rrf{\uvk}p(\uv,\uv')
&=&
\rrf{\uvk}\sum_{l=0}^{l_{\rm max}}\sum_{m=-l}^l{\rm g}^lY_{lm}(\uv)Y_{lm}^*(\uv')
\\
&=&
\sum_{l=0}^{l_{\rm max}}{\rm g}^l\sum_{m'=-l}^l\sum_{m''=-l}^l
e^{-im'\va_{\uvk}}e^{im''\va_{\uvk}}
\\
&\times&
\left(\sum_{m=-l}^l(-1)^{m+m''}d_{m'm}^l(\theta_{\uvk})d_{mm''}^l(\theta_{\uvk})\right)Y_{lm'}(\uv)Y_{lm''}^*(\uv')
\\
&=&
\sum_{l=0}^{l_{\rm max}}\sum_{m=-l}^l{\rm g}^lY_{lm}(\uv)Y_{lm}^*(\uv')=
p(\uv,\uv').
\ea
It is also possible to directly show $\rrf{\uvk}\uv\cdot\uv'=\uv\cdot\uv'$ (see \ref{Wignerpyramid}). Now, we view (\ref{modes:phieq}) in the reference frame whose $z$-axis is rotated to the direction of $\uvk$. By the reverse rotation, we can rewrite (\ref{modes:phieq}) as
\begin{equation}\fl
\left(1-\frac{\mu_i}{\nu}\right)\Phi_{\nu}^m(\uv_i)\approx
\varpi\sum_{l'=0}^{l_{\rm max}}\sum_{m'=-l'}^{l'}{\rm g}^{l'}Y_{l'm'}(\uv_i)
\sum_{i'=1}^{2N}w_{i'}\int_0^{2\pi}Y_{l'm'}^*(\uv_{i'})\Phi_{\nu}^m(\uv_{i'})\,d\va'.
\label{homo1d}
\end{equation}

Let us introduce $g_l^m(\nu)$ as
\begin{equation}\fl
\sum_{i'=1}^{2N}w_{i'}\int_0^{2\pi}Y_{l'm'}^*(\uv_{i'})\Phi_{\nu}^m(\uv_{i'})\,d\va'=
\delta_{mm'}(-1)^m\sqrt{(2l'+1)\pi}g_{l'}^m(\nu),
\label{intYPhi}
\end{equation}
where
\be
g_l^m(\nu)=
(-1)^m\sqrt{\frac{(l-m)!}{(l+m)!}}\sum_{i=1}^{2N}w_i
\phi^m(\nu,\mu_i)\left(1-\mu_i^2\right)^{|m|/2}P_l^m(\mu_i).
\ee
Noting that $P_l^{-m}(\mu_i)=(-1)^m[(l-m)!/(l+m)!]P_l^m(\mu_i)$ and 
$P_m^m(\mu_i)=(-1)^m(2m-1)!!(1-\mu_i^2)^{m/2}$ ($m\ge0$), we obtain
\be
g_m^m(\nu)=\frac{(2m-1)!!}{\sqrt{(2m)!}}=\frac{\sqrt{(2m)!}}{2^mm!},\quad
g_m^{-m}(\nu)=(-1)^mg_m^m(\nu)
\ee
for $m\ge0$. Equation (\ref{homo1d}) is written as
\be\fl
\left(1-\frac{\mu_i}{\nu}\right)\phi^m(\nu,\mu_i)\left(1-\mu_i^2\right)^{|m|/2}
\approx
\frac{\varpi}{2}(-1)^m\sum_{l'=|m|}^{l_{\rm max}}(2l'+1){\rm g}^{l'}
\sqrt{\frac{(l'-m)!}{(l'+m)!}}P_{l'}^m(\mu_i)g_{l'}^m(\nu).
\ee

Let us define
\be
p_l^m(\mu)=
(-1)^m\sqrt{\frac{(l-m)!}{(l+m)!}}P_l^m(\mu)\left(1-\mu^2\right)^{-|m|/2}.
\ee
We have
\be
p_m^m(\mu)=\frac{(2m-1)!!}{\sqrt{(2m)!}}=\frac{\sqrt{(2m)!}}{2^mm!},\quad
p_m^{-m}(\mu)=(-1)^mp_m^m(\mu)
\ee
for $m\ge0$. Moreover from $(l-m+1)P_{l+1}^m(\mu)=(2l+1)\mu P_l^m(\mu)-(l+m)P_{l-1}^m(\mu)$,
\be
\sqrt{(l+1)^2-m^2}p_{l+1}^m(\mu)=(2l+1)\mu p_l^m(\mu)-\sqrt{l^2-m^2}p_{l-1}^m(\mu)
\ee
for $l\ge|m|+1$, and
\be
p_{|m|+1}^m(\mu)=\sqrt{2|m|+1}\mu p_{|m|}^m(\mu).
\ee
For sufficiently large $N$,
\be
\sum_{i=1}^{2N}w_ip_l^m(\mu_i)p_{l'}^m(\mu_i)\left(1-\mu_i^2\right)^{|m|}=
\frac{2}{2l+1}\delta_{ll'}.
\ee
Thus,
\begin{equation}
\left(\nu-\mu_i\right)\phi^m(\nu,\mu_i)\approx
\frac{\varpi\nu}{2}\sum_{l'=|m|}^{l_{\rm max}}(2l'+1){\rm g}^{l'}
p_{l'}^m(\mu_i)g_{l'}^m(\nu).
\label{eigenmode1}
\end{equation}
We note that
\be
g_l^m(\nu)=\sum_{i=1}^{2N}w_i\phi^m(\nu,\mu_i)p_l^m(\mu_i)\left(1-\mu_i^2\right)^{|m|}.
\ee
Assuming that $N$ is sufficiently large, the following recurrence relations hold:
\be
\sqrt{(l+1)^2-m^2}g_{l+1}^m(\nu)+\sqrt{l^2-m^2}g_{l-1}^m(\nu)=\nu h_lg_l^m(\nu)
\ee
for $l\ge|m|+1$, and
\be
\sqrt{2|m|+1}g_{|m|+1}^m(\nu)=\nu h_{|m|}g_{|m|}^m(\nu),
\ee
where
\be
h_l=\cases{
(2l+1)\left(1-\varpi{\rm g}^l\right),
&$0\le l\le l_{\rm max}$,
\\
2l+1,&$l>l_{\rm max}$.
}
\ee
Indeed, $g_l^m(\nu)$ are the normalized Chandrasekhar polynomials \cite{Garcia-Siewert89,Garcia-Siewert90}. For numerical calculation, $g_l^m(\nu)$ can be computed using the recurrence relations. To compute $g_l^m(\nu)$ for $\nu>1$, we define \cite{Garcia-Siewert90}
\be
\tilde{g}_l^m(\nu)=\frac{g_{l+1}^m(\nu)}{g_l^m(\nu)}.
\ee
Then we have
\be
\tilde{g}_{l-1}^m(\nu)=
\frac{\sqrt{l^2-m^2}}{\nu h_l-\sqrt{(l+1)^2-m^2}\tilde{g}_l^m(\nu)}
\ee
for $l=|m|+1,|m|+2,\dots$. We can set $\tilde{g}_l^m(\nu)=0$ when $l$ is sufficiently larger than $l_{\rm max}$. Using $\tilde{g}_l^m(\nu)$, we obtain
\be
g_{l+1}^m(\nu)=\tilde{g}_l^m(\nu)g_l^m(\nu),
\quad l=|m|,|m|+1,\dots.
\ee

From (\ref{eigenmode1}),
\begin{equation}
\phi^m(\nu,\mu_i)\approx
\frac{\varpi\nu}{2}\frac{g^m(\nu,\mu_i)}{\nu-\mu_i},
\label{modes:phidef}
\end{equation}
where
\begin{equation}
g^m(\nu,\mu_i)=
\sum_{l'=|m|}^{l_{\rm max}}(2l'+1){\rm g}^{l'}p_{l'}^m(\mu_i)g_{l'}^m(\nu).
\label{modes:defgm}
\end{equation}
We note that $g^{-m}(\nu,\mu_i)=g^m(\nu,\mu_i)$. If $\nu\neq\uv_i\cdot\uvk$, we have
\be
\phi^m\left(\nu,\uv_i\cdot\uvk(\nu,\vv{q})\right)\approx
\frac{\varpi\nu}{2}\frac{g^m\left(\nu,\uv_i\cdot\uvk(\nu,\vv{q})\right)}{\nu-\uv_i\cdot\uvk(\nu,\vv{q})}.
\ee
Thus,
\begin{equation}\fl
\rrf{\uvk(\nu,\vv{q})}\Phi_{\nu}^m(\uv_i)\approx
\frac{(-1)^m\varpi\nu}{\nu-\uv_i\cdot\uvk(\nu,\vv{q})}\sum_{l=|m|}^{l_{\rm max}}\sqrt{(2l+1)\pi}{\rm g}^lg_l^m(\nu)
\left(\rrf{\uvk(\nu,\vv{q})}Y_{lm}(\uv_i)\right).
\label{modes:modes}
\end{equation}

Hereafter, we assume that $N$ is sufficiently large and the equality numerically holds in (\ref{homo1d}). The above eigenmodes satisfy the orthogonality relation.

\begin{lem}[Machida-Das \cite{Machida-Das22}]
\label{modes:lem:orth}
We have
\be\fl
\sum_{i=1}^{2N}w_i\mu_i\int_0^{2\pi}
\left(\rrf{\uvk(\nu,\vv{q})}\Phi_{\nu}^m(\uv_i)\right)
\left(\rrf{\uvk(\nu',\vv{q})}\Phi_{\nu'}^{m'*}(\uv_i)\right)\,d\va=
2\pi\hat{k}_z(\nu q)\mathcal{N}^m(\nu)\delta_{\nu\nu'}\delta_{mm'},
\ee
where $\mathcal{N}^m(\nu)$ is a positive constant which depends on $m,\nu$.
\end{lem}

The factor $\mathcal{N}^m(\nu)$ in Lemma \ref{modes:lem:orth} satisfies $\mathcal{N}^m(-\nu)=-\mathcal{N}^m(\nu)$ because $g^m(-\nu,-\mu_i)=g^m(\nu,\mu_i)$. We have
\ba
\mathcal{N}^m(\nu)
&=&
\sum_{i=1}^{2N}w_i\mu_i\phi^m(\nu,\mu_i)^2\left(1-\mu_i^2\right)^{|m|}
\\
&=&
\pi(\varpi\nu)^2\sum_{i=1}^{2N}
\frac{w_i\mu_i}{(\nu-\mu_i)^2}\left[\sum_{l=|m|}^{l_{\rm max}}
\sqrt{2l+1}{\rm g}^lY_{lm}(\mu_i,0)g_l^m(\nu)\right]^2.
\ea

Let us calculate eigenvalues. Equation (\ref{homo1d}) can be rewritten as ($i=1,\dots,2N$, $-l_{\rm max}\le m\le l_{\rm max}$)
\be\fl
\left(1-\frac{\mu_i}{\nu}\right)\phi^m(\nu,\mu_i)
=
\frac{\varpi}{2}\sum_{l'=|m|}^{l_{\rm max}}{\rm g}^{l'}(2l'+1)p_{l'}^m(\mu_i)
\sum_{i'=1}^{2N}w_{i'}p_{l'}^m(\mu_{i'})\phi^m(\nu,\mu_{i'}).
\ee
Hence for $i=1,\dots,N$,
\ba
\left(1-\frac{\mu_i}{\nu}\right)\phi^m(\nu,\mu_i)
&=
\frac{\varpi}{2}\sum_{l'=|m|}^{l_{\rm max}}{\rm g}^{l'}(2l'+1)p_{l'}^m(\mu_i)
\\
&\times
\sum_{i'=1}^Nw_{i'}p_{l'}^m(\mu_{i'})
\left[\phi^m(\nu,\mu_{i'})+(-1)^{l'+m}\phi^m(\nu,-\mu_{i'})\right].
\\
\left(1+\frac{\mu_i}{\nu}\right)\phi^m(\nu,-\mu_i)
&=
\frac{\varpi}{2}\sum_{l'=|m|}^{l_{\rm max}}{\rm g}^{l'}(2l'+1)p_{l'}^m(\mu_i)
\\
&\times
\sum_{i'=1}^Nw_{i'}p_{l'}^m(\mu_{i'})
\left[(-1)^{l'+m}\phi^m(\nu,\mu_{i'})+\phi^m(\nu,-\mu_{i'})\right].
\ea
We arrive at the following matrix-vector equation.
\ba
\left(I_N-\frac{1}{\nu}\Xi\right)\vv{\Phi}_+^m(\nu)
&=
\frac{\varpi}{2}\left[W_+^m\vv{\Phi}_+^m(\nu)+W_-^m\vv{\Phi}_-^m(\nu)\right],
\\
\left(I_N+\frac{1}{\nu}\Xi\right)\vv{\Phi}_-^m(\nu)
&=
\frac{\varpi}{2}\left[W_-^m\vv{\Phi}_+^m(\nu)+W_+^m\vv{\Phi}_-^m(\nu)\right],
\ea
where $I_N$ is the $N$-dimensional identity matrix. Matrix $\Xi$ and vectors $\vv{\Phi}_{\pm}^m$ are defined as 
$\Xi=\mathop{\mathrm{diag}}(\mu_1\cdots\mu_N)$, 
$\{\vv{\Phi}_{\pm}^m(\nu)\}_i=\phi^m(\nu,\pm\mu_i)$ ($i=1,\dots,N$). Elements of the matrices $W_{\pm}^m$ are given by
\ba\fl
\{W_{\pm}^m\}_{ij}=
w_j\sum_{l=|m|}^{l_{\rm max}}(2l+1){\rm g}^lp_l^m(\pm\mu_i)p_l^m(\mu_j)
\\
\fl=
w_j\sum_{l=|m|}^{l_{\rm max}}(2l+1){\rm g}^l\frac{(l-m)!}{(l+m)!}
P_l^m(\pm\mu_i)P_l^m(\mu_j)
\left(1-\mu_i^2\right)^{-|m|/2}\left(1-\mu_j^2\right)^{-|m|/2}.
\ea
The fact $\vv{\Phi}_{\pm}^m(-\nu)=\vv{\Phi}_{\mp}^m(\nu)$ implies that $-\nu$ is an eigenvalue if $\nu$ is an eigenvalue.

According to \cite{Barichello11,Siewert00}, we introduce
\be
\vv{U}^m(\nu)=\vv{\Phi}_+^m(\nu)+\vv{\Phi}_-^m(\nu),\quad
\vv{V}^m(\nu)=\vv{\Phi}_+^m(\nu)-\vv{\Phi}_-^m(\nu).
\ee
By adding and subtracting two equations we obtain
\ba
&
\vv{U}^m(\nu)-\frac{1}{\nu}\Xi\vv{V}^m(\nu)=\frac{\varpi}{2}(W_+^m+W_-^m)\vv{U}^m(\nu),
\\
&
\vv{V}^m(\nu)-\frac{1}{\nu}\Xi\vv{U}^m(\nu)=\frac{\varpi}{2}(W_+^m-W_-^m)\vv{V}^m(\nu).
\ea
Hence we obtain \cite{Barichello-Siewert00,Siewert-Wright99}
\begin{equation}
E_-^mE_+^m\Xi\vv{U}^m=\frac{1}{\nu^2}\Xi\vv{U}^m,
\label{evalsys}
\end{equation}
where
\be
E_{\pm}^m=\left[I_N-\frac{\varpi}{2}(W_+^m\pm W_-^m)\right]\Xi^{-1}.
\ee

\section{Energy density}
\label{dIspart}

Let us write
\be\fl
\widetilde{I}(\vv{q},z,\uv_i)=
\mu_t^2\sum_{m=-l_{\rm max}}^{l_{\rm max}}\sum_{n=1}^Na_n^m(\vv{q})
w_{i_0}\mu_{i_0}\left(\rrf{\uvk(\nu_n,\vv{q})}\Phi_{\nu_n}^*(\uv_{i_0})\right)
\left(\rrf{\uvk(\nu_n,\vv{q})}\Phi_{\nu_n}^m(\uv_i)\right)e^{-\hat{k}_z(\nu_nq)z/\nu_n}
\ee
for $z>0$, $0\le i\le 2N$, $\va\in[0,2\pi)$ with unknown coefficients $a_n^m(\vv{q})$. On the boundary, we have
\be
\sum_{m=-l_{\rm max}}^{l_{\rm max}}\sum_{n=1}^Na_n^m(\vv{q})w_{i_0}\mu_{i_0}
\left(\rrf{\uvk(\nu_n,\vv{q})}\Phi_{\nu_n}^*(\uv_{i_0})\right)
\left(\rrf{\uvk(\nu_n,\vv{q})}\Phi_{\nu_n}^m(\uv_i)\right)=
\delta_{ii_0}\delta(\va-\va_0).
\ee
Let us multiply $\rrf{\uvk(\nu_n,\vv{q})}\Phi_{\nu_n}^m(\uv_{i_0})$ and integrate over $\uv_{i_0}$. Using Lemma \ref{modes:lem:orth}, we obtain
\be
a_n^m(\vv{q})=\frac{1}{2\pi\hat{k}_z(\nu_nq)\mathcal{N}^m(\nu_n)}.
\ee
Hence,
\begin{eqnarray}
\widetilde{I}(\vv{q},z,\uv_i)
&=&
\frac{\mu_t^2}{2\pi}\sum_{m=-l_{\rm max}}^{l_{\rm max}}\sum_{n=1}^N
\frac{w_{i_0}\mu_{i_0}}{\hat{k}_z(\nu_nq)\mathcal{N}^m(\nu_n)}
\nonumber \\
&\times&
\left(\rrf{\uvk_+}\Phi_{\nu_n}^{m*}(\uv_{i_0})\right)
\left(\rrf{\uvk_+}\Phi_{\nu_n}^m(\uv_i)\right)e^{-\hat{k}_z(\nu_nq)z/\nu_n}.
\label{solinhalf}
\end{eqnarray}

\subsection{Pencil beam}
\label{density}

Let us put
\be
i_0=N.
\ee
Since $\vv{\omega}_N$ can be replaced by $\vv{0}$ for sufficiently large $N$ ($\vv{\omega}_N\approx\vv{0}$), we have
\ba
&
\rrf{\uvk(\nu_n,\vv{q})}\Phi_{\nu_n}^{m*}(\uv_{i_0})
\\
&=
\frac{(-1)^m\varpi\nu_n}{\nu_n-\uv_{i_0}\cdot\uvk(\nu_n,\vv{q})}
\sum_{l=|m|}^{l_{\rm max}}\sqrt{(2l+1)\pi}{\rm g}^lg_l^m(\nu_n)
\left(\rrf{\uvk(\nu_n,\vv{q})}Y_{lm}^*(\uv_{i_0})\right)
\\
&=
\frac{(-1)^m\varpi\nu_n}{\nu_n-\hat{k}_z(\nu_nq)}
\sum_{l=|m|}^{l_{\rm max}}\frac{2l+1}{2}{\rm g}^lg_l^m(\nu_n)d_{0m}^l[i\tau(\nu_nq)],
\ea
where we used for $\hvv{z}={^t}(0,0,1)$,
\be
Y_{lm}(\uv_N)\approx Y_{lm}(\hvv{z})=\sqrt{\frac{2l+1}{4\pi}}\delta_{m0},
\ee
and
\be
\rrf{\uvk_{\pm}}Y_{lm}(\hvv{z})=
\sqrt{\frac{2l+1}{4\pi}}d_{0m}^l[i\tau(\nu_nq)].
\ee
Equation (\ref{solinhalf}) becomes
\ba
\widetilde{I}(\vv{q},z,\uv_i)
&=
\frac{\mu_t^2}{4\pi}\sum_{m=-l_{\rm max}}^{l_{\rm max}}\sum_{n=1}^N
\frac{w_N\mu_Ne^{-\hat{k}_z(\nu_nq)z/\nu_n}}{\hat{k}_z(\nu_nq)\mathcal{N}^m(\nu_n)}
\left(\rrf{\uvk_+}\Phi_{\nu_n}^m(\uv_i)\right)
\\
&\times
\frac{(-1)^m\varpi\nu_n}{\nu_n-\hat{k}_z(\nu_nq)}
\sum_{l=|m|}^{l_{\rm max}}(2l+1){\rm g}^lg_l^m(\nu_n)d_{0m}^l[i\tau(\nu_nq)].
\ea

We note that
\be
\sum_{i=1}^{2N}w_i\int_0^{2\pi}\rrf{\uvk}\Phi_{\nu_n}^m(\uv_i)\,d\va\approx
\sum_{i=1}^{2N}w_i\int_0^{2\pi}\Phi_{\nu_n}^m(\uv_i)\,d\va=2\pi\delta_{m0}.
\ee
The energy density is given by
\be
u(\vv{r})=
\frac{1}{(2\pi)^2}\int_{\Rm^2}e^{i\vv{q}\cdot\vv{\rho}}\sum_{i=1}^{2N}w_i
\int_0^{2\pi}\widetilde{I}(\vv{q},z,\uv_i)\,d\va d\vv{q}.
\ee
We have
\begin{equation}
u(\vv{r})=
\frac{1}{4\pi}\varpi\mu_t^2\int_0^{\infty}J_0(q\rho)F(q,z)q\,dq,
\label{u1}
\end{equation}
where
\be\fl
F(q,z)=
\sum_{n=1}^N
\frac{w_N\mu_N\nu_ne^{-\hat{k}_z(\nu_n^0q)z/\nu_n^0}}{\hat{k}_z(\nu_nq)\mathcal{N}^0(\nu_n)\left(\nu_n-\hat{k}_z(\nu_nq)\right)}
\sum_{l=0}^{l_{\rm max}}(2l+1){\rm g}^lg_l^0(\nu_n)d_{00}^l[i\tau(\nu_nq)].
\ee
Here, we used the Hansen-Bessel formula (\ref{asol:HansenBessel}):
\begin{equation}
J_0(x)=\frac{1}{2\pi}\int_0^{2\pi}e^{ix\cos\va}\,d\va
\label{asol:HansenBessel}
\end{equation}
with $J_0(x)$ ($x\ge0$) the Bessel function of order $0$.

\subsection{Isotropic source}
\label{isos}

Instead of (\ref{sourceg}), let us suppose
\begin{equation}
g(\vv{\rho},\uv_i)=\delta(\vv{\rho}).
\label{sourcegiso}
\end{equation}
We note that
\be
\sum_{i_0=1}^{2N}w_{i_0}\int_0^{2\pi}\mu_{i_0}\rrf{\uvk(\nu,\vv{q})}\Phi_{\nu}^{0*}(\uv_{i_0})\,d\va_0=
2\pi\hat{k}_z(\nu q)\nu\left(1-\varpi\right).
\ee
By integrating (\ref{solinhalf}) with respect to $\uv_{i_0}$,
\be\fl
\sum_{i_0=1}^{2N}\int_0^{2\pi}\widetilde{I}(\vv{q},z,\uv_i)\,d\va_0=
\left(1-\varpi\right)
\mu_t^2\sum_{m=-l_{\rm max}}^{l_{\rm max}}\sum_{n=1}^N
\frac{\nu_ne^{-\hat{k}_z(\nu_nq)z/\nu_n}}{\mathcal{N}^0(\nu_n)}
\left(\rrf{\uvk_+}\Phi_{\nu_n}^m(\uv_i)\right).
\ee
The energy density is obtained as
\begin{equation}
u(\vv{r})=
\left(1-\varpi\right)\mu_t^2\int_0^{\infty}J_0(q\rho)F_{\rm iso}(q,z)q\,dq,
\label{isou}
\end{equation}
where
\be
F_{\rm iso}(q,z)=
\sum_{n=1}^N\frac{\nu_n^0}{\mathcal{N}^0(\nu_n^0)}
e^{-\hat{k}_z(\nu_n^0q)z/\nu_n^0}.
\ee

\section{Numerical calculation}
\label{dnum}

\subsection{Nonlinear scattering for the pencil beam}

Suppose $\vv{\rho}\neq\vv{0}$. Let us return to (\ref{intro:rte0}) from the radiative transport equation (\ref{intro:rte}) with dimensionless variables. Recovering the subscript $*$, (\ref{u1}) is written as
\begin{equation}
u_*(\vv{r}_*)=
\frac{1}{4\pi}\varpi\mu_t^2\int_0^{\infty}J_0(q\rho_*)F(q,z_*)q\,dq.
\label{J2}
\end{equation}
We note that $q$ is already dimensionless, and $\rho_*=\mu_t\rho$, $z_*=\mu_tz$.

For the purpose of numerical calculation, we use the asymptotic expression of the Bessel function:
\ba
&\fl
J_0(x)=\sqrt{\frac{2}{\pi x}}
\\
&\fl\times
\left[
\left(1-\frac{9}{128x^2}+O(x^{-4})\right)\cos\left(x-\frac{\pi}{4}\right)+
\left(\frac{1}{8x}-\frac{75}{1024x^3}+O(x^{-5})\right)\sin\left(x-\frac{\pi}{4}\right)
\right].
\ea
We define
\ba
&\fl
d(q,\rho_*)=
qJ_0(q\rho_*)-\sqrt{\frac{2q}{\pi\rho_*}}
\\
&\fl\times
\left[
\left(1-\frac{9}{128(q\rho_*)^2}\right)\cos\left(q\rho_*-\frac{\pi}{4}\right)+
\left(\frac{1}{8q\rho_*}-\frac{75}{1024(q\rho_*)^3}\right)\sin\left(q\rho_*-\frac{\pi}{4}\right)
\right].
\ea
Let us set
\begin{equation}
a=\frac{\pi}{4\rho_*}.
\label{smalla}
\end{equation}
We have
\begin{eqnarray}
u_*(\vv{r}_*)
&=
\frac{\varpi\mu_t^2}{4\pi}\int_0^aqJ_0(q\rho_*)F(q,z_*)\,dq+
\frac{\varpi\mu_t^2}{4\pi}\int_a^{\infty}d(q,\rho_*)F(q,z_*)\,dq
\nonumber \\
&+
\frac{\varpi\mu_t^2}{(2\pi)^{3/2}\sqrt{\rho_*}}\int_0^{\infty}\left[f_1(s,\vv{r}_*)+f_2(s,\vv{r}_*)\right]\,ds,
\label{final_U}
\end{eqnarray}
where $s=q-a$,
\begin{eqnarray}
&\fl
f_1(s,\vv{r}_*)=
\sqrt{s+\frac{\pi}{4\rho_*}}\left(
1-\frac{9}{128(s\rho_*+\frac{\pi}{4})^2}\right)
F\left(s+\frac{\pi}{4\rho_*},z_*\right)\cos(s\rho_*),
\nonumber \\
&\fl
f_2(s,\vv{r}_*)=
\sqrt{s+\frac{\pi}{4\rho_*}}\left(
\frac{1}{8(s\rho_*+\frac{\pi}{4})}-\frac{75}{1024(s\rho_*+\frac{\pi}{4})^3}
\right)
F\left(s+\frac{\pi}{4\rho_*},z_*\right)\sin(s\rho_*).
\label{f1f2}
\end{eqnarray}
The integrals $\int_0^{\infty}F_1(q)\,dq$ and $\int_0^{\infty}F_2(q)\,dq$ can be evaluated by the double-exponential formula \cite{Ooura-Mori91,Ooura-Mori99,Ogata05}. Define
\be
\phi(\tau)=\frac{\tau}{1-e^{-6\sinh{\tau}}}
\ee
with
\be
\phi'(\tau)=\frac{1-(1+6\tau\cosh{\tau})e^{-6\sinh{\tau}}}{\left(1-e^{-6\sinh{\tau}}\right)^2}.
\ee
With the approximation of the double-exponential formula, we can write
\ba
&\fl
\int_0^{\infty}f_1(s,\vv{r}_*)\,ds\approx
\frac{\pi}{\rho_*}\sum_{k=-N_k}^{N_k}f_1\left(\frac{\pi}{h\rho_*}\phi\left(kh+\frac{h}{2}\right),\vv{r}_*\right)
\phi'\left(kh+\frac{h}{2}\right),
\\
&\fl
\int_0^{\infty}f_2(s,\vv{r}_*)\,ds\approx
\frac{\pi}{\rho_*}\sum_{k=-N_k}^{N_k}f_2\left(\frac{\pi}{h\rho_*}\phi(kh),\vv{r}_*\right)
\phi'(kh),
\ea
where $N_k>0$ is an integer and $h$ is a mesh size.

\begin{figure}[ht]
\begin{center}
\includegraphics[width=0.8\textwidth]{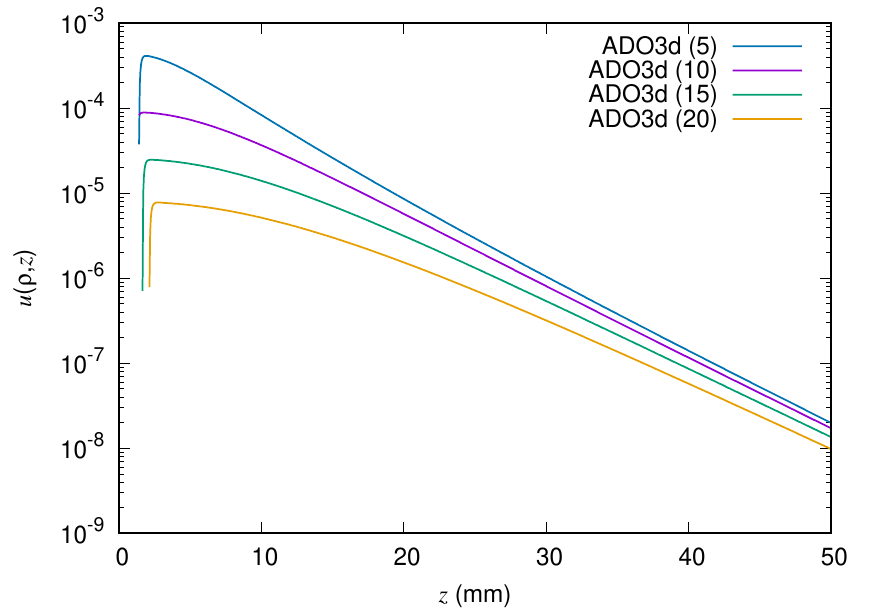}
\end{center}
\caption{
From the top, $\rho=5,10,15,20\,{\rm mm}$. Parameters are set as $\mu_a=0.01\,{\rm mm}^{-1}$, $\mu_s=10\,{\rm mm}^{-1}$, ${\rm g}=0.9$, $l_{\rm max}=9$, $N=9$.
}
\label{fig_ado1}
\end{figure}

Results are shown in Fig.~\ref{fig_ado1}.

\subsection{Nonlinear scattering for the isotropic source}
\label{withmc}

\begin{figure}[ht]
\begin{center}
\includegraphics[width=0.8\textwidth]{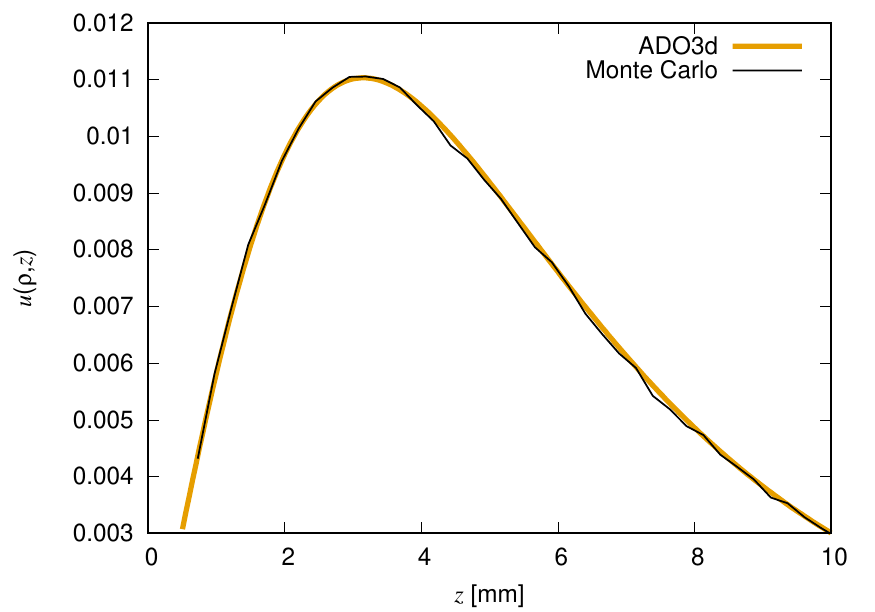}
\end{center}
\caption{
The energy density for the isotropic source. The solution from the proposed method, which is denoted by ADO3d, is compared with Monte Carlo simulation. Parameters are set as $\rho=5\,{\rm mm}$, $\mu_a=0.01\,{\rm mm}^{-1}$, $\mu_s=10\,{\rm mm}^{-1}$, ${\rm g}=0.9$, $l_{\rm max}=9$, $N=9$.
}
\label{fig_case3}
\end{figure}

The case of $l_{\rm max}=9$ for the isotropic source (\ref{sourcegiso}) is considered. In Fig.~\ref{fig_case3}, the result is compared to Monte Carlo simulation \cite{MarkelMC} . The Monte Carlo code assumes the Henyey-Greenstein model \cite{Henyey-Greenstein41} for the scattering phase function (i.e., $l_{\rm max}\to\infty$) and $10^8$ photons were launched.

\subsection{Linear scattering for the isotropic source}

We consider the isotropic source (\ref{sourcegiso}). Here, we set
\be
l_{\rm max}=1.
\ee
Equation (\ref{isou}) is numerically computed as
\ba
&&\fl
u_*(\vv{r}_*)=
(1-\varpi)\mu_t^2\int_0^aqJ_0(q\rho_*)F_{\rm iso}(q,z_*)\,dq+
(1-\varpi)\mu_t^2\int_a^{\infty}d(q,\rho_*)F_{\rm iso}(q,z_*)\,dq
\\
&&+
\frac{2(1-\varpi)\mu_t^2}{\sqrt{2\pi\rho_*}}\int_0^{\infty}\left[f_{{\rm iso},1}(s,\vv{r}_*)+f_{{\rm iso},2}(s,\vv{r}_*)\right]\,ds,
\ea
where $s=q-a$ with $a$ defined in (\ref{smalla}). The functions $f_{{\rm iso},1},f_{{\rm iso},2}$ are given by (\ref{f1f2}) when $F$ is replaced by $F_{\rm iso}$.

\begin{figure}[ht]
\begin{center}
\includegraphics[width=0.8\textwidth]{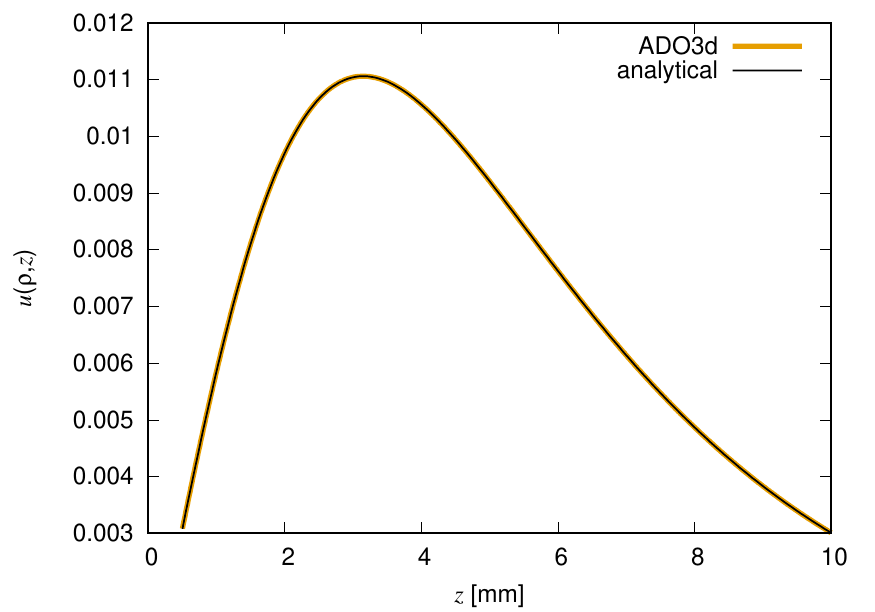}
\end{center}
\caption{
The energy density for the isotropic source. The solution from the proposed method, which is denoted by ADO3d, is compared to the analytical solution. Parameters are set as $\rho=5\,{\rm mm}$, $\mu_a=0.01\,{\rm mm}^{-1}$, $\mu_s=10\,{\rm mm}^{-1}$, ${\rm g}=0.9$, $l_{\rm max}=1$, $N=9$.
}
\label{fig_case2}
\end{figure}

In this case, an analytical solution is available (see \ref{singular}). In Fig.~\ref{fig_case2}, results are shown together with the curve from the analytical solution (see \ref{singular}).

\section{Concluding remarks}
\label{concl}

As an alternative approach, the problem for the half space with the vacuum boundary condition can be solved by the expansion of singular eigenfunctions with spherical harmonics \cite{Machida-etal10,Machida14,Panasyuk06}. This spherical-harmonic approach suffers from divergence even for isotropic scattering. In the discrete-ordinate approach developed in this paper, increasing $N$ does not cause divergence.

In this paper, the unidirectional source (\ref{sourceg}) and isotropic source (\ref{sourcegiso}) were considered. In general, the solution $I(\vv{r},\uv_i)$ for a source $g(\vv{\rho},\uv_i)$ can be obtained as
\be
I(\vv{r},\uv_i)=
\sum_{i_0=1}^N\int_0^{2\pi}\int_{\Rm^2}G(\vv{\rho}-\vv{\rho}',z;\uv_i,\uv_{i_0})g(\vv{\rho}',\uv_{i_0})\,d\vv{\rho}'d\va_0.
\ee
Here,
\be
G(\vv{\rho}-\vv{\rho}',z;\uv_i,\uv_{i_0})=
\frac{1}{(2\pi)^2}\int_{\Rm^2}e^{i\vv{q}\cdot(\vv{\rho}-\vv{\rho}')}
\widetilde{I}(\vv{q},z,\uv_i)\,d\vv{q},
\ee
where $\widetilde{I}(\vv{q},z,\uv_i)\,$ is given in (\ref{solinhalf}).

\section*{Acknowledgments}
The author acknowledges support from JSPS KAKENHI Grant Number JP17K05572 and JST PRESTO Grant Number JPMJPR2027.

\appendix

\section{Analytically continued Wigner $d$-matrices}
\label{Wignerpyramid}

The initial term is given by
\be
d^0_{00}[i\tau(x)]=1.
\ee
Moreover we define
\be
d^1_{00}=\sqrt{1+x^2},\quad
d^1_{01}=i\frac{|x|}{\sqrt{2}},\quad
d^1_{1,\pm1}=\frac{1\pm\sqrt{1+x^2}}{2}.
\ee
Analytically continued Wigner $d$-matrices can be computed with a pyramid scheme using recurrence relations \cite{Blanco97}. We note that
\be
d^{l}_{mm'}=d^{l}_{-m',-m}=(-1)^{m+m'}d^{l}_{-m,-m'}=(-1)^{m+m'}d^{l}_{m'm}.
\ee

For each value of $l$ ($2\le l\le l_{\rm max}$), we first compute $d^l_{mm'}[i\tau(x)]$ ($m=0,\dots,l-2;m'=-m,\dots,m$) according to
\ba
\fl
d^l_{mm'}=
\frac{l(2l-1)}{\sqrt{(l^2-m^2)(l^2-{m'}^2)}}
\\
\times
\left[\left(d^1_{00}-\frac{mm'}{l(l-1)}\right)d^{l-1}_{mm'}
-\frac{\sqrt{\left[(l-1)^2-m^2\right]\left[(l-1)^2-{m'}^2\right]}}{(l-1)(2l-1)}
d^{l-2}_{mm'}\right].
\ea
We obtain $d^l_{ll}[i\tau(x)]$ and $d^l_{l-1,l-1}[i\tau(x)]$ as
\be
d^l_{ll}=d^1_{11}d^{l-1}_{l-1,l-1},
\quad
d^l_{l-1,l-1}=(ld^1_{00}-l+1)d^{l-1}_{l-1,l-1},
\ee
and $d^l_{lm'}[\tau(x)]$ $(m'=l-1,\dots,-l)$ as
\be
d^l_{lm'}=-i\sqrt{\frac{l+m'+1}{l-m'}}
\sqrt{\left|\frac{d^1_{1-1}}{d^1_{11}}\right|}d^{l}_{l,m'+1}.
\ee
With the relation
\be
d^l_{l-1,m'}=-i\frac{ld^1_{00}-m'}{ld^1_{00}-m'-1}
\sqrt{\frac{l+m'+1}{l-m'}}\sqrt{\left|\frac{d^1_{1-1}}{d^1_{11}}\right|}
d^{l}_{l-1,m'+1},
\ee
we have $d^l_{l-1,m'}[i\tau(x)]$ ($m'=l-2,\dots,1-l$).

Let us show $\rrf{\uvk}\uv\cdot\uv'=\uv\cdot\uv'$. We have
\ba
\rrf{\uvk}\uv\cdot\uv'
&=&
\rrf{\uvk}\left[\sin\theta\sin\theta'\cos(\va-\va')+\cos\theta\cos\theta'\right]
\\
&=&
\frac{4\pi}{3}\rrf{\uvk}\left[Y_{11}(\uv)Y_{11}^*(\uv')+Y_{11}^*(\uv)Y_{11}(\uv')+Y_{10}(\uv)Y_{10}(\uv')\right],
\ea
where we used $(-1)^mY_{l,-m}(\uv)=Y_{lm}^*(\uv)$. Thus,
\ba
&&\fl
\rrf{\uvk}\uv\cdot\uv'=
-\frac{4\pi}{3}\left(e^{i\va_{\uvk}}d_{1,-1}^1(\theta_{\uvk})Y_{1,-1}(\uv)+d_{01}^1(\theta_{\uvk})Y_{10}(\uv)+e^{-i\va_{\uvk}}d_{11}^1(\theta_{\uvk})Y_{11}(\uv)\right)
\\
&&\fl
\times\left(e^{i\va_{\uvk}}d_{11}^1(\theta_{\uvk})Y_{1,-1}(\uv')-d_{01}^1(\theta_{\uvk})Y_{10}(\uv')+e^{-i\va_{\uvk}}d_{1,-1}^1(\theta_{\uvk})Y_{11}(\uv')\right)
\\
&&\fl
-\frac{4\pi}{3}\left(e^{i\va_{\uvk}}d_{11}^1(\theta_{\uvk})Y_{1,-1}(\uv)-d_{01}^1(\theta_{\uvk})Y_{10}(\uv)+e^{-i\va_{\uvk}}d_{1,-1}^1(\theta_{\uvk})Y_{11}(\uv)\right)
\\
&&\fl
\times\left(e^{i\va_{\uvk}}d_{1,-1}^1(\theta_{\uvk})Y_{1,-1}(\uv')+d_{01}^1(\theta_{\uvk})Y_{10}(\uv')+e^{-i\va_{\uvk}}d_{11}^1(\theta_{\uvk})Y_{11}(\uv')\right)
\\
&&\fl
\frac{4\pi}{3}\left(e^{i\va_{\uvk}}d_{01}^1(\theta_{\uvk})Y_{1,-1}(\uv)+d_{00}^1(\theta_{\uvk})Y_{10}(\uv)-e^{-i\va_{\uvk}}d_{01}^1(\theta_{\uvk})Y_{11}(\uv)\right)
\\
&&\fl
\times\left(e^{i\va_{\uvk}}d_{01}^1(\theta_{\uvk})Y_{1,-1}(\uv')+d_{00}^1(\theta_{\uvk})Y_{10}(\uv')-e^{-i\va_{\uvk}}d_{01}^1(\theta_{\uvk})Y_{11}(\uv')\right).
\ea
Noting that $d_{00}^1,d_{1,\pm1}^1$ are real and $d_{01}^1$ is pure imaginary, we obtain
\ba
&&\fl
\rrf{\uvk}\uv\cdot\uv'=
\frac{8\pi}{3}\Biggl\{
\left[\left(d_{01}^1(\theta_{\uvk})\right)^2-2d_{11}^1(\theta_{\uvk})d_{1,-1}^1(\theta_{\uvk})\right]\Re e^{2i\va_{\uvk}}Y_{11}^*(\uv)Y_{11}^*(\uv')
\\
&&\fl+
\left[\left(d_{11}^1(\theta_{\uvk})\right)^2+\left(d_{01}^1(\theta_{\uvk})\right)^2+\left(d_{1,-1}^1(\theta_{\uvk})\right)^2\right]\Re Y_{11}^*(\uv)Y_{11}(\uv')
\\
&&\fl+
\left[\left(d_{01}^1(\theta_{\uvk})\right)^2+\frac{1}{2}\left(d_{00}^1(\theta_{\uvk})\right)^2\right]Y_{10}(\uv)Y_{10}(\uv')
\\
&&\fl-
i\left|d_{01}^1(\theta_{\uvk})\right|\left[-d_{11}^1(\theta_{\uvk})+d_{00}^1(\theta_{\uvk})+d_{1,-1}^1(\theta_{\uvk})\right]\Re e^{i\va_{\uvk}}Y_{10}(\uv)Y_{11}^*(\uv')
\\
&&\fl+
i\left|d_{01}^1(\theta_{\uvk})\right|\left[d_{11}^1(\theta_{\uvk})-d_{00}^1(\theta_{\uvk})-d_{1,-1}^1(\theta_{\uvk})\right]\Re e^{i\va_{\uvk}}Y_{11}^*(\uv)Y_{10}(\uv')\Biggr\}.
\ea
By substituting $d_{00}^1,d_{01}^1,d_{1,\pm1}^1$, we find
\be
\rrf{\uvk}\uv\cdot\uv'=
\frac{8\pi}{3}\left\{\Re Y_{11}^*(\uv)Y_{11}(\uv')+
\frac{1}{2}Y_{10}(\uv)Y_{10}(\uv')\right\}=
\uv\cdot\uv'.
\ee

\section{Singular eigenfunctions in three dimensions}
\label{singular}

\subsection{Radiative transport equation}
\label{appen:rte}

We consider the following radiative transport equation.
\begin{equation}\fl
\uv\cdot\nabla I(\vv{r},\uv)+\mu_tI(\vv{r},\uv)=
\mu_s\int_{\Sm^2}p(\uv,\uv')I(\vv{r},\uv')\,d\uv',\quad
(\vv{r},\uv)\in\Omega\times\Sm^2
\label{asol:rte1}
\end{equation}
with constants $\mu_s,\mu_t$ ($0<\mu_s<\mu_t$). Note that $p(\uv,\uv')$ in (\ref{phasefunc}) satisfies
\be
\int_{\Sm^2}p(\uv,\uv')\,d\uv'=1,\quad
\int_{\Sm^2}(\uv\cdot\uv')p(\uv,\uv')\,d\uv'={\rm g}.
\ee
The unit vector $\uv\in\Sm^2$ is introduced as
\be
\uv=\left(\begin{array}{c}\vv{\omega}\\ \mu\end{array}\right),
\qquad
\vv{\omega}=\left(\begin{array}{c}
\sqrt{1-\mu^2}\cos\va \\ \sqrt{1-\mu^2}\sin\va
\end{array}\right)
\ee
for $-1\le\mu\le1$, $0\le\va<2\pi$. We impose the boundary condition as
\be
I(\vv{r},\uv)=\delta(\vv{\rho})\delta(\uv-\uv''),
\quad\vv{r}\in\pp\Omega,\quad 0<\mu\le1,\quad 0\le\va<2\pi,
\ee
where $\delta(\uv-\uv'')=\delta(\mu-\mu'')\delta(\va-\va'')$ with $\mu'',\va''$ the polar and azimuthal angles of $\uv''$.

By dividing both sides of (\ref{asol:rte1}) by $\mu_t$, we obtain
\begin{equation}
\cases{
\uv\cdot\nabla_*I_*(\vv{r}_*,\uv)+I_*(\vv{r}_*,\uv)=
\varpi\int_{\Sm^2}p(\uv,\uv')I_*(\vv{r}_*,\uv')\,d\uv',&
\\
\qquad\vv{r}\in\Omega,\quad \uv\in\Sm^2,&
\\
I_*(\vv{r}_*,\uv)=\mu_t^2\delta(\vv{\rho}_*)\delta(\uv-\uv''),&
\\
\qquad\vv{r}\in\pp\Omega,\quad 0<\mu\le1,\quad 0\le\va<2\pi.&
}
\label{asol:rte2}
\end{equation}
Hereafter we will take the unit of length to be $1/\mu_t$ and drop $*$.

We assume the specific intensity of the form
\begin{equation}
\widetilde{I}_{\nu}^m(\vv{r},\uv;\vv{q})=
\left(\rrf{\uvk}\Phi_{\nu}^m(\uv)\right)e^{-\uvk\cdot\vv{r}/\nu},
\label{appen:modes:ansatz}
\end{equation}
where $\uvk$ is given in (\ref{defveck}) and
\begin{equation}
\Phi_{\nu}^m(\uv)
=\phi^m(\nu,\mu)\left(1-\mu^2\right)^{|m|/2}e^{im\va}.
\label{appen:modes:ansatz2}
\end{equation}
These $\Phi_{\nu}^m(\uv),\phi^m(\nu,\mu)$ are called singular eigenfunctions. Note that $\phi^m(\nu,\mu)$ in (\ref{appen:modes:ansatz2}) is different from $\phi^m(\nu,\mu_i)$ in (\ref{modes:Phiphi}). We will see below that the former contains generalized functions. We normalize $\phi^m$ as
\begin{equation}\fl
\frac{1}{2\pi}\int_{\Sm^2}\rrf{\uvk}\phi^m(\nu,\mu)\left(1-\mu^2\right)^{|m|}\,d\uv=
\int_{-1}^1\phi^m(\nu,\mu)\left(1-\mu^2\right)^{|m|}\,d\mu=
1.
\label{appen:modes:normalization}
\end{equation}
In the laboratory frame ($\uvk=\hvv{z}$), (\ref{appen:modes:ansatz}) reduces to the form used in \cite{McCormick-Kuscer66}. Similar to (\ref{modes:homoeq}), we will determine elementary solutions $\widetilde{I}_{\nu}^m(\vv{r},\uv;\vv{q})$ in (\ref{appen:modes:ansatz}) so that they satisfy the following homogeneous equation:
\begin{equation}\fl
\mu\frac{\pp}{\pp z}\widetilde{I}_{\nu}^m(\vv{q},z,\uv')+
(1+i\vv{\omega}\cdot\vv{q})\widetilde{I}_{\nu}^m(\vv{q},z,\uv')=
\varpi\int_{\Sm^2}p(\uv,\uv')\widetilde{I}_{\nu}^m(\vv{q},z,\uv')\,d\uv'.
\label{appen:homoeq}
\end{equation}

By plugging (\ref{appen:modes:ansatz}) into (\ref{appen:homoeq}), we obtain
\begin{eqnarray}
\fl
\rrf{\uvk}\left(1-\frac{\mu}{\nu}\right)\phi^m(\nu,\mu)
\left(1-\mu^2\right)^{|m|/2}e^{im\va}
\nonumber \\
\fl=
\varpi\int_{\Sm^2}p\left(\rrf{\uvk}\uv,\rrf{\uvk}\uv'\right)
\rrf{\uvk}\phi^m(\nu,\mu')\left(1-{\mu'}^2\right)^{|m|/2}e^{im\va'}\,d\uv',
\label{appen:rte:RTEansatz}
\end{eqnarray}
where we used $p(\uv,\uv')=p(\rrf{\uvk}\uv,\rrf{\uvk}\uv')$. Following \cite{McCormick-Kuscer66}, we express $p(\uv,\uv')$ in (\ref{phasefunc}) as
\be\fl
p(\uv,\uv')=
\sum_{l=0}^{l_{\rm max}}\sum_{m=-l}^l{\rm g}^l\frac{2l+1}{4\pi}
\left(1-\mu^2\right)^{|m|/2}
\left(1-{\mu'}^2\right)^{|m|/2}p_l^m(\mu)p_l^m(\mu')e^{im(\va-\va')}.
\ee
The right-hand side of (\ref{appen:rte:RTEansatz}) is calculated as
\ba
\fl
\mbox{RHS}=
2\pi\varpi\Theta\left(l_{\rm max}-|m|\right)
\left[\rrf{\uvk}\left(1-\mu^2\right)^{|m|/2}e^{im\va}\right]
\\
\fl\times
\sum_{l'=|m|}^{l_{\rm max}}{\rm g}^{l'}\frac{2l'+1}{4\pi}
\left[\rrf{\uvk}p_{l'}^m(\mu)\right]\int_{-1}^1p_{l'}^m(\mu')\phi^m(\nu,\mu')
\left(1-{\mu'}^2\right)^{|m|}\,d\mu',
\ea
where the step function $\Theta(\cdot)$ was defined as $\Theta(x)=1$ for $x\ge0$ and $=0$ for $x<0$.  Hence,
\begin{equation}
\fl
\rrf{\uvk}\left(\nu-\mu\right)\phi^m(\nu,\mu)=
2\pi\varpi\nu\Theta\left(l_{\rm max}-|m|\right)
\sum_{l'=|m|}^{l_{\rm max}}{\rm g}^{l'}\frac{2l'+1}{4\pi}
\rrf{\uvk}p_{l'}^m(\mu)g_{l'}^m(\nu),
\label{appen:phiisgiven}
\end{equation}
where
\be
g_l^m(\nu)=\int_{-1}^1\phi^m(\nu,\mu)p_l^m(\mu)\left(1-\mu^2\right)^{|m|}\,d\mu.
\ee
These $g_l^m$ are the normalized Chandrasekhar polynomials \cite{Garcia-Siewert89,Garcia-Siewert90}. Similar polynomials were also considered by Mika \cite{Mika61} and McCormick-Ku\v{s}\v{c}er \cite{McCormick-Kuscer66}. Since the right-hand side of (\ref{appen:phiisgiven}) is zero for $|m|>l_{\rm max}$ and then $\phi^m=0$, hereafter we suppose
\be
0\le|m|\le l_{\rm max}.
\ee

Similar to (\ref{modes:phidef}), $\phi^m$ is obtained as
\begin{equation}
\phi^m(\nu,\mu)=
\frac{\varpi\nu}{2}\mathcal{P}\frac{g^m(\nu,\mu)}{\nu-\mu}+
\lambda^m(\nu)(1-\nu^2)^{-|m|}\delta(\nu-\mu),
\label{appen:eigenfunction}
\end{equation}
where $\mathcal{P}$ denotes the Cauchy principal value, $g^m(\nu,\mu)$ is given in (\ref{modes:defgm}), and $\lambda^m(\nu)$ is defined below.

By multiplying $\left(1-\mu^2\right)^{|m|}$ and integrating over $\uv$, (\ref{appen:eigenfunction}) becomes
\be
1=\frac{\varpi\nu}{2}\pint_{-1}^1\frac{g^m(\nu,\mu)}{\nu-\mu}
\left(1-\mu^2\right)^{|m|}\,d\mu+
\int_{-1}^1\lambda^m(\nu)\delta(\nu-\mu)\,d\mu.
\ee

For $\nu\in(-1,1)$ we obtain
\be
\lambda^m(\nu)=
1-\frac{\varpi\nu}{2}\pint_{-1}^1\frac{g^m(\nu,\mu)}{\nu-\mu}
\left(1-\mu^2\right)^{|m|}\,d\mu.
\ee
Note that $\lambda^{-m}(\nu)=\lambda^m(\nu)$ and hence $\phi^{-m}(\nu,\mu)=\phi^m(\nu,\mu)$.

Let us define
\be
\Lambda^m(z)=
1-\frac{\varpi z}{2}\int_{-1}^1\frac{g^m(z,\mu)}{z-\mu}
\left(1-\mu^2\right)^{|m|}\,d\mu,
\ee
where $z\in\Cm$.  Eigenvalues $\nu\notin[-1,1]$ are solutions to 
\begin{equation}
\Lambda^m(\nu)=0.
\label{appen:rootsearch}
\end{equation}
We write these discrete eigenvalues as $\pm\nu_j^m$ ($\nu_0^m>\nu_1^m>\cdots>\nu_{M-1}^m>1$). Note that $\nu_j^{-m}=\nu_j^m$. The number of discrete eigenvalues $M^m$ depends on $|m|$ and we have \cite{Mika61,McCormick-Kuscer66} $M^m\le N-|m|+1$. For $\nu\in(-1,1)$, we have the continuous spectrum.

Now, let us express $\widetilde{I}$ as
\ba
&&\fl
\widetilde{I}(\vv{q},z,\uv)=
\mu_t^2\sum_{m=-l_{\rm max}}^{l_{\rm max}}\sum_{j=1}^Ma_j^m(\vv{q})
\mu''\left(\rrf{\uvk(\nu_j^m,\vv{q})}\Phi_{\nu_j^m}^*(\uv'')\right)
\left(\rrf{\uvk(\nu_j^m,\vv{q})}\Phi_{\nu_j^m}^m(\uv)\right)
e^{-\hat{k}_z(\nu_j^mq)z/\nu_j^m}
\\
&&\fl+
\mu_t^2\sum_{m=-l_{\rm max}}^{l_{\rm max}}\int_0^1a(\nu,\vv{q})
\mu''\left(\rrf{\uvk(\nu,\vv{q})}\Phi_{\nu}^*(\uv'')\right)
\left(\rrf{\uvk(\nu,\vv{q})}\Phi_{\nu}^m(\uv)\right)
e^{-\hat{k}_z(\nu q)z/\nu}\,d\nu
\ea
for $z>0$, $\uv\in\Sm^2$. From the boundary, we have
\ba
&&\fl
\sum_{m=-l_{\rm max}}^{l_{\rm max}}\sum_{j=1}^Na_j^m(\vv{q})
\left(\rrf{\uvk(\nu_j^m,\vv{q})}\Phi_{\nu_j^m}^*(\uv'')\right)
\left(\rrf{\uvk(\nu_j^m,\vv{q})}\Phi_{\nu_j^m}^m(\uv)\right)
\\
&&\fl+
\sum_{m=-l_{\rm max}}^{l_{\rm max}}\int_0^1a(\nu,\vv{q})
\left(\rrf{\uvk(\nu,\vv{q})}\Phi_{\nu}^*(\uv'')\right)
\left(\rrf{\uvk(\nu,\vv{q})}\Phi_{\nu}^m(\uv)\right)\,d\nu
=
\delta(\mu-\mu_0)\delta(\va-\va_0).
\ea
Let us multiply $\rrf{\uvk(\xi,\vv{q})}\Phi_{\xi}^m(\uv'')$ ($\xi=\nu_j^m$, $j=1,\dots,M$ or $\xi=\nu\in(0,1)$) and integrate over $\uv_0$. Using the orthogonality relations \cite{Machida14}, we obtain
\be
a_j^m(\vv{q})=\frac{1}{2\pi\hat{k}_z(\nu_j^mq)\mathcal{N}^m(\nu_j^m)},\quad
a(\nu,\vv{q})=\frac{1}{2\pi\hat{k}_z(\nu q)\mathcal{N}^m(\nu)},
\ee
where
\be
\mathcal{N}^m(\nu)=\int_{-1}^1\mu\phi^m(\nu,\mu)^2\left(1-\mu^2\right)^{|m|}\,d\mu.
\ee
Hence,
\begin{eqnarray}
\fl
\widetilde{I}(\vv{q},z,\uv)=
\frac{\mu_t^2}{2\pi}\sum_{m=-l_{\rm max}}^{l_{\rm max}}\Biggl[
\sum_{j=0}^{M^m-1}
\frac{\mu''e^{-\hat{k}_z(\nu_j^mq)z/\nu_j^m}}{\hat{k}_z(\nu_j^mq)\mathcal{N}^m(\nu_j^m)}
\left(\rrf{\uvk(\nu_j^m,\vv{q})}\Phi_{\nu_j^m}^{m*}(\uv'')\right)
\left(\rrf{\uvk(\nu_j^m,\vv{q})}\Phi_{\nu_j^m}^m(\uv)\right)
\nonumber \\
\fl+
\int_0^1\frac{\mu''}{\hat{k}_z(\nu q)\mathcal{N}^m(\nu)}
\left(\rrf{\uvk(\nu,\vv{q})}\Phi_{\nu}^{m*}(\uv'')\right)
\left(\rrf{\uvk(\nu,\vv{q})}\Phi_{\nu}^m(\uv)\right)
e^{-\hat{k}_z(\nu q)z/\nu}\,d\nu
\Biggr].
\label{asol:solinhalf}
\end{eqnarray}

\begin{rmk}
If singular eigenfunctions in (\ref{asol:solinhalf}) are expanded by spherical harmonics, we arrive at the solution in the half space given in \cite{Schotland-Markel07}.
\end{rmk}

Using (\ref{asol:solinhalf}), the energy density is calculated as
\ba
u(\vv{r})
&=&
\int_{\Sm^2}I(\vv{r},\uv)\,d\uv
\\
&=&
\frac{1}{(2\pi)^2}\int_{\Rm^2}e^{i\vv{q}\cdot\vv{\rho}}\int_{\Sm^2}\widetilde{I}(\vv{q},z,\uv)\,d\uv d\vv{q}
\\
&=&
\frac{\mu_t^2}{(2\pi)^2}\int_{\Rm^2}e^{i\vv{q}\cdot\vv{\rho}}
\Biggl[
\sum_{j=0}^{M^0-1}\frac{\mu''}{\hat{k}_z(\nu_j^0q)\mathcal{N}^0(\nu_j^0)}
\left(\rrf{\uvk(\nu_j^0,\vv{q})}\Phi_{\nu_j^0}^{0*}(\uv'')\right)
e^{-\hat{k}_z(\nu_j^0q)z/\nu_j^0}
\\
&+&
\int_0^1\frac{\mu''}{\hat{k}_z(\nu q)\mathcal{N}^0(\nu)}
\left(\rrf{\uvk(\nu,\vv{q})}\Phi_{\nu}^{0*}(\uv'')\right)
e^{-\hat{k}_z(\nu q)z/\nu}\,d\nu
\Biggr]\,d\vv{q}.
\ea

Let us calculate the energy density for an isotropic source $\delta(\vv{\rho})$. For this purpose, we note that
\ba
&&
\int_{\Sm^2}\mu''\rrf{\uvk(\nu,\vv{q})}\Phi_{\nu}^{m*}(\uv'')\,d\uv''
\\
&=&
\int_{\Sm^2}\left(\irrf{\uvk(\nu,\vv{q})}\mu''\right)\Phi_{\nu}^{m*}(\uv'')\,d\uv''
\\
&=&
\int_0^{2\pi}\int_{-1}^1\left(\hat{k}_z(\nu q)\mu''-i|\nu q|\sqrt{1-(\mu'')^2}\cos\va''\right)
\left[\phi^m(\nu,\mu'')\left(1-(\mu'')^2\right)^{|m|/2}e^{-im\va''}\right]
\,d\mu''d\va''
\\
&=&
\int_0^{2\pi}\int_{-1}^1\hat{k}_z(\nu q)\mu''
\left[\phi^m(\nu,\mu'')\left(1-(\mu'')^2\right)^{|m|/2}e^{-im\va''}\right]
\,d\mu''d\va''
\\
&-&
i|\nu q|\int_0^{2\pi}\int_{-1}^1\sqrt{1-(\mu'')^2}\cos\va''
\left[\phi^m(\nu,\mu'')\left(1-(\mu'')^2\right)^{|m|/2}e^{-im\va''}\right]
\,d\mu''d\va''.
\ea
Thus,
\ba
&&
\int_{\Sm^2}\mu''\rrf{\uvk(\nu,\vv{q})}\Phi_{\nu}^{m*}(\uv'')\,d\uv''
\\
&=&
2\pi\hat{k}_z(\nu q)\delta_{m0}\int_{-1}^1\mu''\phi^0(\nu,\mu'')\,d\mu''
\\
&-&
i\pi|\nu q|\left[\delta_{m1}\int_{-1}^1\left(1-(\mu'')^2\right)\phi^1(\nu,\mu'')\,d\mu''+
\delta_{m,-1}\int_{-1}^1\left(1-(\mu'')^2\right)\phi^{-1}(\nu,\mu'')\,d\mu''\right].
\ea
In particular,
\be
\int_{\Sm^2}\mu''\rrf{\uvk(\nu,\vv{q})}\Phi_{\nu}^{0*}(\uv'')\,d\uv''=
2\pi\hat{k}_z(\nu q)\int_{-1}^1\mu''\phi^0(\nu,\mu'')\,d\mu''.
\ee
Noticing
\be
\left(1-\frac{\mu}{\nu}\right)\phi^0(\nu,\mu)=\frac{\varpi}{2}
\sum_{l=0}^{l_{\rm max}}{\rm g}^lP_l(\mu)\int_{-1}^1P_l(\mu')\phi^0(\nu,\mu')\,d\mu',
\ee
we have
\begin{eqnarray}
&
\int_{\Sm^2}\mu''\rrf{\uvk(\nu,\vv{q})}\Phi_{\nu}^{0*}(\uv'')\,d\uv''
\nonumber \\
&=
2\pi\hat{k}_z(\nu q)\int_{-1}^1\left[\nu\phi^0(\nu,\mu'')-\frac{\varpi\nu}{2}
\sum_{l=0}^{l_{\rm max}}{\rm g}^lP_l(\mu'')\int_{-1}^1P_l(\mu')\phi^0(\nu,\mu')\,d\mu'\right]\,d\mu''
\nonumber \\
&=
2\pi\hat{k}_z(\nu q)\nu\left(1-\varpi\right).
\label{asol:int23}
\end{eqnarray}
We obtain
\begin{eqnarray}
&&\fl
u(\vv{r})=
\int_{\Sm^2}I(\vv{r},\uv)\,d\uv=
\frac{1}{(2\pi)^2}\int_{\Rm^2}e^{i\vv{q}\cdot\vv{\rho}}\int_{\Sm^2}\widetilde{I}(\vv{q},z,\uv)\,d\uv d\vv{q}
\nonumber \\
&&\fl=
\mu_t^2(1-\varpi)\int_0^{\infty}qJ_0(q\rho)
\nonumber \\
&&\fl\times
\left[\sum_{j=0}^{M^0-1}\frac{\nu_j^0}{\mathcal{N}^0(\nu_j^0)}e^{-\hat{k}_z(\nu_j^0q)z/\nu_j^0}+
\int_0^1\frac{\nu}{\mathcal{N}^0(\nu)}
e^{-\hat{k}_z(\nu q)z/\nu}\,d\nu
\right]\,dq.
\label{analyticalu}
\end{eqnarray}

\subsection{Linear scattering for the isotropic source}

If $l_{\rm max}=1$, then $M^0=1$. We can write $\nu_0=\nu_0^0>1$. We note that $g_0^0(\nu_0)=1$, $g_1^0(\nu)=(1-\varpi)\nu$. Let us consider \cite{Case60,Case-Zweifel}
\ba
\Lambda(z)
&=&
1-\frac{\varpi z}{2}\int_{-1}^1\frac{g^0(z,\mu)}{z-\mu}\,d\mu
\\
&=&
1-\frac{\varpi z}{2}\sum_{l'=0}^1(2l'+1){\rm g}^{l'}g_{l'}^0(z)\int_{-1}^1\frac{P_{l'}(\mu)}{z-\mu}\,d\mu
\\
&=&
1-\frac{\varpi z}{2}\left[\int_{-1}^1\frac{1}{z-\mu}\,d\mu+
3{\rm g}(1-\varpi)z\int_{-1}^1\frac{\mu}{z-\mu}\,d\mu\right]
\\
&=&
\left(1+3{\rm g}(1-\varpi)z^2\right)\Lambda_{l_{\rm max}=0}(z)-3{\rm g}(1-\varpi)^2z^2.
\ea
The function $\Lambda_{l_{\rm max}=0}(z)$ for isotropic case is obtained as
\be
\Lambda_{l_{\rm max}=0}(z)=1-\varpi z\tanh^{-1}\frac{1}{z}.
\ee
We have
\be
\lim_{\epsilon\to0^{\pm}}\Lambda_{l_{\rm max}=0}(\nu+i\epsilon)=
\lambda(\nu)\pm\frac{i\pi\varpi\nu}{2},\quad\nu\in(-1,1),
\ee
where
\be
\lambda(\nu)=1-\varpi\nu\tanh^{-1}\nu.
\ee
Hence,
\ba
\Lambda(z)
&=&
\left(1+3{\rm g}(1-\varpi)z^2\right)\left(1-\varpi z\tanh^{-1}\frac{1}{z}\right)-3{\rm g}(1-\varpi)^2z^2
\\
&=&
1+3{\rm g}\varpi(1-\varpi)z^2
-\left(1+3{\rm g}(1-\varpi)z^2\right)\varpi z\tanh^{-1}\frac{1}{z}.
\ea
We obtain $\nu_0$ as a positive root of $\Lambda(z)$:
\be
\Lambda(\nu_0)=0.
\ee
The normalization factor is given by \cite{Case-Zweifel}
\ba
\mathcal{N}^0(\nu_0)
&=&
\int_{-1}^1\mu\phi(\nu_0,\mu)^2\,d\mu=
\frac{\varpi\nu_0^2}{2}g^0(\nu_0,\nu_0)\left.\frac{d\Lambda(z)}{dz}\right|_{z=\nu_0},
\\
\mathcal{N}^0(\nu)
&=&
\int_{-1}^1\mu\phi(\nu,\mu)^2\,d\mu=
\nu\left(\lim_{\epsilon\to0^+}\Lambda(\nu+i\epsilon)\right)
\left(\lim_{\epsilon\to0^-}\Lambda(\nu+i\epsilon)\right),
\ea
where $g^0(\nu_0,\nu_0)=1+3{\rm g}(1-\varpi)\nu_0^2$. We obtain
\ba
&&\fl
\frac{d\Lambda(z)}{dz}=
6{\rm g}\varpi(1-\varpi)z
-\varpi\left(1+9{\rm g}(1-\varpi)z^2\right)\tanh^{-1}\frac{1}{z}
+\varpi\left(1+3{\rm g}(1-\varpi)z^2\right)\frac{z}{z^2-1}
\\
&&\fl=
9{\rm g}\varpi(1-\varpi)z
-\varpi\left(1+9{\rm g}(1-\varpi)z^2\right)\tanh^{-1}\frac{1}{z}
+\varpi\left(1+3{\rm g}(1-\varpi)\right)\frac{z}{z^2-1},
\ea
and
\ba
\lim_{\epsilon\to0^{\pm}}\Lambda(\nu+i\epsilon)
&=&
\left(1+3{\rm g}(1-\varpi)\nu^2\right)
\lim_{\epsilon\to0^{\pm}}\Lambda_{l_{\rm max}=0}(\nu+i\epsilon)-
3{\rm g}(1-\varpi)^2\nu^2
\\
&=&
\left(1+3{\rm g}(1-\varpi)\nu^2\right)
\left(\lambda(\nu)\pm\frac{i\pi\varpi\nu}{2}\right)-
3{\rm g}(1-\varpi)^2\nu^2
\\
&=&
\left(1+3{\rm g}(1-\varpi)\nu^2\right)\lambda(\nu)
-3{\rm g}(1-\varpi)^2\nu^2
\pm\frac{i\pi\varpi\nu}{2}\left(1+3{\rm g}(1-\varpi)\nu^2\right).
\ea

The double-exponential formula can be used to evaluate the integral over $q$ in (\ref{analyticalu}). We have
\be
u(\vv{r})=
\left(1-\varpi\right)\mu_t^2\int_0^{\infty}J_0(q\rho)F_a(q,z)q\,dq,
\ee
where
\be
F_a(q,z)=
\frac{\nu_0e^{-\hat{k}_z(\nu_0q)z/\nu_0}}{\mathcal{N}^0(\nu_0)}+
\int_0^1\frac{\nu e^{-\hat{k}_z(\nu q)z/\nu}}{\mathcal{N}^0(\nu)}\,d\nu.
\ee
Thus,
\ba
&&\fl
u_*(\vv{r}_*)=
(1-\varpi)\mu_t^2\int_0^aqJ_0(q\rho_*)F_a(q,z_*)\,dq+
(1-\varpi)\mu_t^2\int_a^{\infty}d(q,\rho_*)F_a(q,z_*)\,dq
\\
&&+
\frac{2(1-\varpi)\mu_t^2}{\sqrt{2\pi\rho_*}}\int_0^{\infty}\left[f_{a,1}(s,\vv{r}_*)+f_{a,2}(s,\vv{r}_*)\right]\,ds,
\ea
where $s=q-a$. Here, the positive constant $a$ is given in (\ref{smalla}) and $f_{a,1},f_{a,2}$ are obtained by using $F_a$ instead of $F$ in (\ref{f1f2}).

\end{document}